# Surface plasmon based sensing with broadband coherent laser pulses

**ALEXANDRE KOLOMENSKII,**[1,*] **ELIZABETH SUROVIC,**[1] **AND HANS SCHUESSLER**[1,2]

[1]*Department of Physics and Astronomy, Texas A&M University, College Station, TX 77843-4242, USA,*
[2]*Science Department, Texas A&M University at Qatar, Doha 23874, Qatar*
*\*Corresponding author: alexandr@physics.tamu.edu*

A broadband detection method with coherent laser pulses exciting surface plasmons is investigated. The intensity and phase characteristics of the reflected light in Kretschmann configuration are calculated. It is shown that by mixing in-plane and out-of-plain polarization components of the incident light, the measurements of the changes of the refractive index of the adjacent media can be efficiently performed by observing the shift of the steep slope of the surface plasmon resonance (SPR) reflectance curve. We have demonstrated that such sensing can be performed by observing the angular shift of the SPR curve as well as the spectral shift. The developed model can account for an arbitrary number of layers and is used to simulate the response when an additional functionalizing layer adjacent to the gold film is used.

## 1. Introduction

The surface plasmon resonance (SPR) has been broadly used for chemical and bio-sensing for several decades. Since surface plasmons (SPs) are propagating along the metal surface and are confined to the proximity of the surface, this effect is sensitive to changes of the dielectric properties of the media close to the metal, supporting plasmonic oscillations, which makes the SPR highly suitable for sensing [1]. The most common arrangement for realization of such sensing employs the Kretschmann configuration [2], however some sensor realizations with the Otto configuration were also proposed [3].

The SPR can be easily realized with CW monochromatic laser radiation, and in particular, laser diodes were commonly used as light sources for sensing purposes [4,5] with the wavelengths fulfilling the necessary requirement that the frequency of the incident light be lower than the plasmon frequency [6]. Remarkable properties of SPs are broadly utilized in nano-photonics [7], biological applications [8,9], and opto-electronics [10]. SPs are efficiently excited when the phase velocity of the trace of the electromagnetic wave incident on the surface matches that of SPs [11], which can be realized at a certain (resonance) incidence angle. The most traditional approach to sensing is to measure the changes of the intensity of the reflected light, when the light's incidence angle is close to resonance, and this approach also works with incoherent light sources [12,13].

The variations of the dielectric constants of the media involved in the SPR lead to changes in the intensity and also in the phase of the reflected light. Therefore, early on it became clear that the phase changes in the vicinity of the resonance and, in particular, interferometric methods for detection of phase shifts in SPR due to refractive index variations could also be used for bio- and chemical sensing [14]. Consequently, the phase changes in the Kretschmann SPR configuration were studied and applied for measurements and imaging of the refractive index changes [15]. Interferometric techniques for detection and imaging of phase variations were also employed for measuring spatial distributions in an array configuration for bio-molecular interaction analysis [16]. The phase effects manifesting themselves in Fano resonances were observed in nanostructures [17] and used for development of on-chip sensors [18]. Phase changes in light interactions with nano-hole arrays exhibiting SP resonances were investigated and utilized for sensing as well [19]. Nano-structures paired with an infrared plasmonic sensing device have been used for multiplexing and remote sensing [20], and plasmonic slit arrays were proposed for applications in high-efficiency polychromatic filtering and ultracompact biosensing [21].

There are several advantages and new sensing modalities when using broadband light. Multi-wavelength detection can be used for broadband absorption measurements of films using the reflected intensity as well as by detecting the phase response [22]. SP resonances with incoherent radiation were used for fluorescent ultra-sensitive detection in bioassays [23] and for hyperspectral-imaging of binding biomolecules in a miniaturized sensor [24].

Recently, there was significant progress in the generation of stable trains of ultrashort laser pulses used to produce spectrally broadband frequency combs [25]. In this respect, it is of interest to analyze the special features and possibilities of the interaction of coherent broadband pulses with systems supporting plasmonic modes and consider their utilization for sensing [26]. The interaction of ultrashort pulses with SP polaritons was used to excite waveguide resonances [27,28]. Multiplexing of SP measuring channels in combination with multi-wavelength spectral analysis was used for simultaneous analysis of several analytes [29]. The design with specially prepared substrates to reduce non-specific binding was employed for accurate label-free biosensing [30].

Several studies employed broadband coherent light for plasmonic excitations with nano-antennas and nano-ring arrays, showing ultrahigh resolution and sensitivity [31,32]. Fabry-Pérot cavities have been used in conjunction with the SPR, and the changes of intensity and phase have been examined over a wavelength interval, demonstrating high sensitivity for biosensing [33]. Especially sharp plasmonic resonances can be realized in the infrared, where broadband detection was also realized [34]. In structures used for SPR sensing, the metallic losses are commonly relatively low, assuring sharp resonances. However, with a special design for photovoltaic applications a high absorption and low reflection were realized [35]. Methods to compensate for the dispersion of SPs and to enable the resonance excitation in a broad spectral range have also been developed [36].

Despite all these advances a detailed consideration of the interaction of broadband coherent laser pulses with surface plasmons and the possibility of sensing with such pulses was lacking. In this work, we theoretically investigate the conversion of femtosecond pulses into SPs at the incidence angles close to the resonant one. We show that the broad bandwidth of the pulse strongly affects the spectra of the generated SPs as well as the reflected pulses. Since the refractive indexes vary for different wavelengths, the angular shift of the resonance point for different wavelengths leads to some smearing of the SPR dip observed in the reflected pulse. We show that by mixing the light components of p- and s- polarizations a steep slope in the reflected light can be observed, wherein the s-polarization component not interacting with SPs can play the role of a reference that can be used to control the degree of asymmetry of the Fano resonance. The response characteristics to variations of the refractive index of the medium adjacent to the metal (gold) film were calculated, and the model can be easily extended to sandwich-like structures with arbitrary number of layers.

## 2. Description of the interaction of light with surface plasmons

We assume that the excitation of SPs is performed in the Kretschmann configuration (Fig. 1) and consider a short laser pulse, which for simplicity and in order to obtain analytical expressions, is assumed to be of Gaussian temporal shape

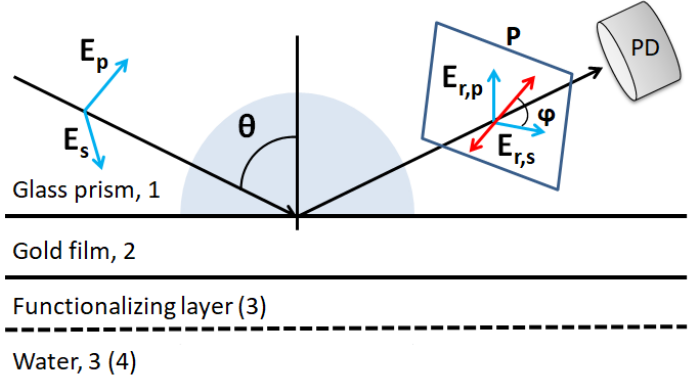

**Fig. 1.** Schematic of the Kretschmann configuration considered. The light is incident from the glass prism (the 1st medium) on the gold film (the 2nd medium) at a certain angle θ, which is bordering a liquid (3rd medium). When an optional functionalizing layer is present, this layer becomes the 3rd medium and the liquid layer is the 4th medium. The p- and s-polarization components of the light electric field, $E_p$ and $E_s$, are shown. After reflection the s- and p-polarization components become $E_{r,p}$ and $E_{r,s}$, and the light passes a polarizer (P) with the polarization orientation shown by the red arrows (at an angle φ to the $E_{r,s}$), and then the light is registered by the photodetector (PD).

$$I(t) = I_0 f(t), \ f(t) = \exp(-t^2/\tau^2), \tag{1}$$

with the duration at FWHM $\tau_0 = 2\sqrt{\ln(2)}\tau$ and with a plane wavefront, which is incident from a dielectric (medium 1) on a metal film (gold, medium 2) at an angle θ, and the metal film is adjacent to a dielectric liquid (water, medium 3). Further, we will also consider the configuration, where an additional functionalizing layer is used, in which case the latter is the 3rd medium and the liquid is then the 4th medium.

The spectrum of the pulse can be described as

$$I(\omega) = I_0 \sqrt{\pi}\tau \exp[-(\omega-\omega_0)^2 \tau^2/4], \tag{2}$$

where $\omega_0 = 2\pi c/\lambda$ is the carrier angular frequency, λ and c are the wavelength and the speed of light in the air, and the frequency ($f=\omega/2\pi$) bandwidth of the pulse is

$$\Delta f = \frac{\sqrt{\ln 2}}{\pi \tau_0}. \tag{3}$$

If the incident light has both in-plane (p) and out-of-plane (s) components of the electric field, then these two polarization components experience reflection with different reflection coefficients:

$$V_{p,s} = \frac{Z_{in,2;p,s} - Z_{1;p,s}}{Z_{in,2;p,s} + Z_{1;p,s}}, \tag{4}$$

where

$$Z_{in,n;p,s} = \frac{Z_{in,n+1;p,s} - iZ_{n;p,s}\tan(k_n d_n)}{Z_{n;p,s} - iZ_{in,n+1;p,s}\tan(k_n d_n)} Z_{n;p,s}, n = 2, 3, \ldots, \tag{5}$$

and

$$k_{z,n} = \sqrt{\varepsilon_n k_0^2 - k_1^2}, \; k_1 = \sqrt{\varepsilon_1} k_0 \sin\theta, \; k_0 = 2\pi/\lambda,$$

$$Z_{n;p} = \frac{k_n}{\varepsilon_n k_0}, \; Z_{n;s} = \frac{k_0}{k_n}, Z_{in,N;p,s} = Z_{N;p,s}, Z_{in,N-1;p,s} = Z_{N-1;p,s}. \quad (6)$$

Here the iterative calculation proceeds starting from the outer layer (N) by obtaining expressions for $Z_{in,n;p,s}$ from known $Z_{in,n+1;p,s}$ for both p and s polarizations for n=N-1,N-2...1. The component of the p-polarized light can excite SPs and therefore will experience strong variations of the intensity and phase in the vicinity of the resonance angle.

The determination of the phase change can be done by interfering the reflected p-component with a reference beam. As such, the light with s-polarization can be used, which does not interact with SPs. By placing a polarizer in front of the photodetector at an angle φ, the intensity of the resultant field found by the summation of the projections of the p- and s-polarization components on the transmission direction of the polarizer is detected, which can be described by relations

$$E_{r,p-s} = \left(E_{s,p} e^{i\psi_1} \sin\varphi + E_{r,s} e^{i\psi_2} \cos\varphi\right) f(\omega), f(\omega) = \sqrt{2\pi}\tau \exp\left[2(\omega-\omega_0)^2/\tau^2\right], \quad (7)$$

$$E_{r,p} = |V_p| E_p, \; E_{r,s} = |V_s| E_s, \; \psi_p = \arg[V_p], \psi_s = \arg[V_s]. \quad (8)$$

In particular cases φ=0° and φ=90° the above expression describes the p- and s-polarization components, respectively. The spectral intensity of the reflected pulse is

$$I_{r,p-s}(\omega) = \frac{c_0 n_1 \varepsilon_0}{2} |E_{r,p-s}(\omega)|^2, \quad (9)$$

where the time dependence of the field of the reflected pulse can be found by the inverse Fourier transform,

$$E_{p-s}(t) = \frac{1}{2\pi} \int E_{p,s} \left[\cos(\omega t) + i\sin(\omega t)\right] d\omega. \quad (10)$$

The total flux of the light on the photodetector is the integral of the intensity over time, i.e.

$$F_{r,p-s} = \int_{-\infty}^{+\infty} I_{r,p-s}(t) \, dt. \quad (11)$$

For sensing applications, when the spectral mode of detection is used, the variations of the spectral intensity of the reflected pulse, $I_{r,p-s}(\omega)$, are measured in response to a change of the refractive index of the liquid layer, Δn. When the angular variations of the reflected light are used for sensing, it is important to know the changes of the SPR curve (in this case, the dependence of the total reflected light flux on the angle) in response to a change of the refractive index of the liquid layer, Δn, and the modified SPR curve can be found as $F(n_{3(4)} + \Delta n)$, where $n_{3(4)} = \sqrt{\varepsilon_{3(4)}}$.

## 3. Results of the calculations

### 3.1 SPR with a monochromatic light

Let us first consider the reflection of a single spectral component for p- and s-polarizations. We consider the Kretschmann configuration (Fig. 1) and in all calculations assume the wavelength of 800nm and a 45nm thick gold film. For p-polarization, the intensity of the reflected light shows a dip at the SPR resonance angle (Fig. 2(a)), while the phase at resonance experiences a stepwise change (Fig. 2(b)) (the dependences for p-polarization are shown by red lines). For s-polarization, the variations of the intensity and the phase for the same angular interval are relatively small and gradual (Fig. 2(a,b), the dependences for s-polarization are shown by blue lines). The inset depicts the s-polarization on a finer scale, showing that the phase variations are indeed rather small. The kinks observed near $\theta_r=61.8°$ are due to the onset of the total internal reflection. Figure 2(c) depicts the density plot of the reflectance versus the angle $\theta$, and wavelength $\lambda$, for a p-polarized light. In the density plot the purple color corresponds to the minimum of the reflection, and the red color shows the reflection close to maximum. A sharp change of color at $\theta_r=61.8°$ signifies the onset of the total internal reflection. For different wavelengths the resonance angle changes, as follows from the deviation of the purple streak across the density plot in Fig.2(c) from the vertical. Thus, both p-and s-components experience the total internal reflection, however only the p-polarization component shows a dip in the reflection due to the interaction with surface plasmons.

Now, we assume that the incident light has equal p- and s-polarization components and is in *a mixed polarization state* with a polarizer set in front of the photodetector with an axis at $\varphi=45°$ with respect to $E_p$ and $E_s$ (Fig. 1). This arrangement is assumed further on, when we talk about the mixed polarization state, unless indicated otherwise. Then, the projections of $E_p$ and $E_s$ on the polarization axis interfere, and the result depends on the relative phase. As one can see, the phases $\psi_p$, $\psi_s$ are not differing that much for angles below the resonance angle $\theta_r=66.2°$, so that the p- and s-polarization component interfere constructively. However, the phases $\psi_p$, $\psi_s$ differ significantly for the angle $\theta$ above the resonance angle, where a destructive interference takes place. Consequently, for the intensity of the reflected light in the mixed polarization state a steep drop of intensity is observed in proximity of the resonance angle (Fig. 3(a)). Figure 3(b) shows the variation of the phase vs. angle of a reflected pulse with mixed polarizations. Since the reflected p-component is strongly suppressed as the angle is increasing and approaches the resonance angle, the phase at resonance gets closer to the phase of the s-component, and then with the further increase of the incidence angle it drops and plateaus at a somewhat higher level.

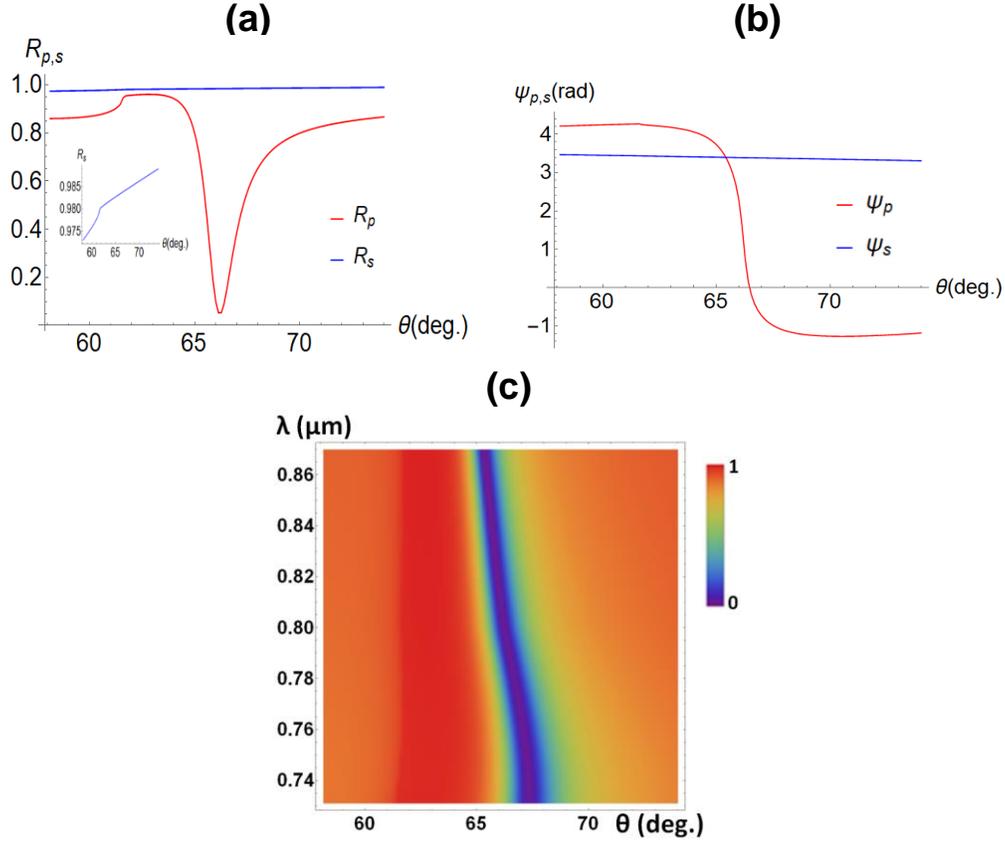

**Fig. 2.** Angular dependences for the intensity of the reflected light of p- and s- polarizations for a wavelength of 800nm. (c). The distribution of the reflected spectral intensity of a pulse near the SPR over wavelength $\lambda$ and the incident angle $\theta$. The visible change of the color at the angle $\theta_c \sim 61.5°$ is due to the onset of the total internal reflection. The purple streak shows the minima of the reflection indicating also the resonance conditions with the angular positions somewhat shifted for different wavelengths; $\theta=66.2°$ corresponds to the resonance angle at the central wavelength 800nm. All the calculations are for a wavelength 800nm, a gold film thickness of 45nm and water as the adjacent (third) medium.

The p-polarization component of light excites an SP wave, and the reflected wave depends on their interference. This reflected wave shows asymmetry of the angular and spectral dependences due to phase variations. When the p- and s-polarizations are mixed, the asymmetry can be strongly enhanced as a result of a transition from constructive to destructive interference of the two components, when the angle or the wavelength is going through the resonance point. Such a behavior is a well-known feature of a Fano resonance [17], where the resonantly interacting wave with abrupt phase variations interferes with another (background) wave that does not experience significant variations of the phase. In our case, the ratio of the two contributions can be changed by changing the orientation of the polarizer in front of the photodetector. Consequently, the degree of the asymmetry can be easily controlled (engineered) in this way.

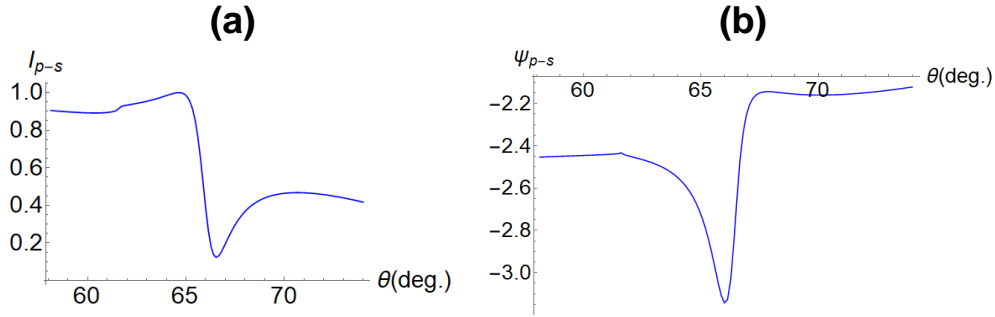

**Fig. 3.** The angular dependences of the reflected light in a mixed polarization state incident at the angle of 66.2° with initially (at the incidence) equal p- and s-polarization components detected with the polarizer set at $\varphi=45°$: (a) The normalized intensity angular dependence. (b) The angular dependence of the phase.

### 3.2 SPR with a broadband optical pulse

We now will consider the interaction with plasmons of a broadband pulse. We performed calculations having in mind pulse parameters of a Rainbow oscillator (Spectra-Physics® Vienna, the former Femtolasers) with typical values for the pulse duration, $\tau_0=7$fs and the central wavelength, 800nm. The data for the media dielectric constants for different wavelength were accepted from the following sources: for BK7 glass ($\varepsilon_1$, [37] ), gold ($\varepsilon_2$, [38]), and water ($\varepsilon_3$, [39]), and the data were interpolated for the wavelength interval from 700 to 900nm.

First, we consider the properties of the SPR for a p-polarized pulse. Figure 4(a) shows the normalized spectral intensity of the initial p-polarized pulse (blue solid line) and of its reflection (black solid line) and also of the reflection, when the same pulse is s-polarized (red solid line).

When the SPR is well pronounced, the losses (internal and radiative) are relatively small, and therefore the observed SPR dip is mainly due to the conversion of the light into SPs. Thus, assuming small losses, the SP spectrum can be approximated as the difference of the spectra of the incidence pulse and the reflected pulse for p-polarization, and it is shown in Fig. 4(a) by a black dashed line.

Figure 4(b) shows the spectral intensity distribution of the wave packets of SPs excited by the laser pulse at different incidence angles. This distribution was found by subtracting the spectrum of the reflected p-polarized pulse from the spectrum of the incident pulse. Figure 4(c) shows in the time domain the shapes of the reflected pulse sent at different incidence angles. The total reflected flux of the pulsed radiation incident on the gold film at different angles, i.e. the SPR curve for the pulsed excitation, is shown in Fig 4(d).

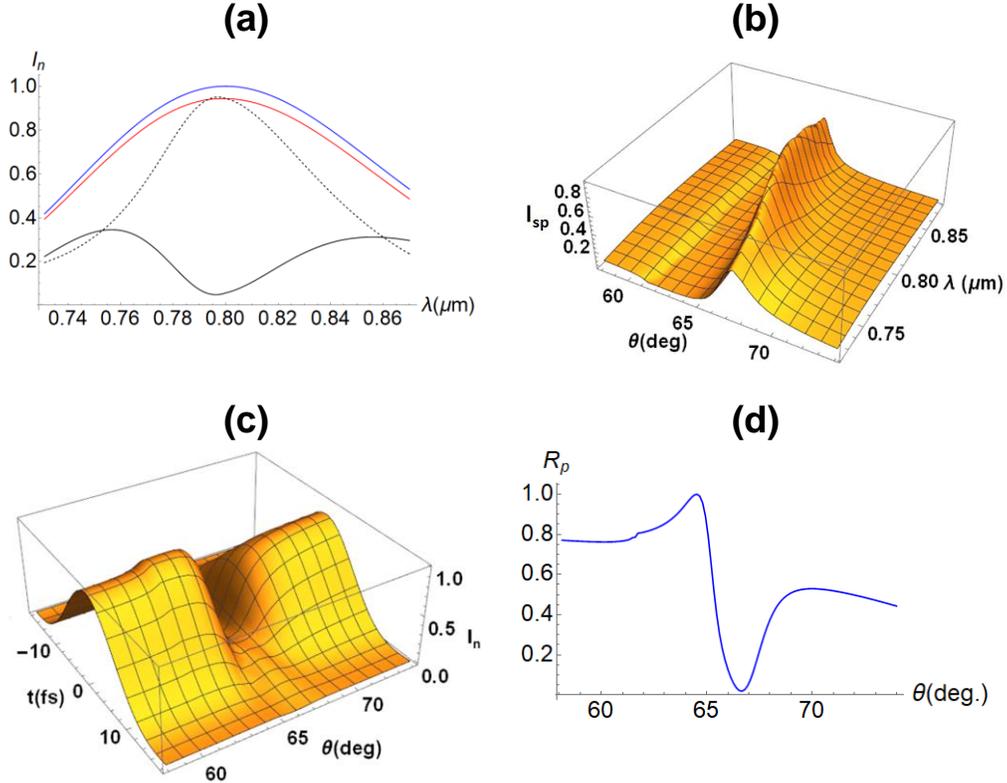

**Fig. 4.** Plots for a 7 fs FWHM pulse with the central wavelength 800nm interacting with SPs: (a) The spectrum of the incident p-polarized or s-polarized pulse (blue line), the spectrum of the reflected p-polarized pulse (black line), the spectrum of the reflected s-polarized pulse (red line), the spectrum of the excited SPs (black dashed line) for an incidence angle 66.2°. (b) The distribution of the spectral intensity of the SPs excited by a p-polarized pulse vs. incidence angle $\theta$ and wavelength $\lambda$. (c) A 3D-plot showing the temporal shapes of the reflected p-polarized pulse at different incidence angles $\theta$. (d) The dependence of the total radiation flux of the reflected p-polarized pulses, sent at different incidence angles.

*3.3 SPR with a broadband optical pulse of mixed p-s-polarizations*

For a pulse of a mixed polarization state the distribution of the spectral intensity over angle and wavelength can be calculated from Eqs. (7-9), and the result is shown in Fig. 5(a). This distribution shows a steep negative slope with increasing wavelength values. The profile of the reflected pulses for the mixed polarization state experiences significant changes in the vicinity of the resonance $\theta_r \sim 66.2°$ (Fig. 5(b)), since the intensity of the reflected pulses is strongly suppressed for $\theta_r$ exceeding the resonance incidence angle.

If the incidence angle is fixed at the resonance angle for a specific wavelength, then the dependence of the spectral intensity distribution on the wavelength also shows a steep negative slope near the resonant wavelength. Figure 5(c) depicts the spectral intensity distribution for the incidence angle $\theta=66.2°$ with the spectral components showing strongly reduced magnitudes for wavelengths above 800nm. The flux detected for different incidence angles exhibits a strongly asymmetric dip (Fig. 5(d)).

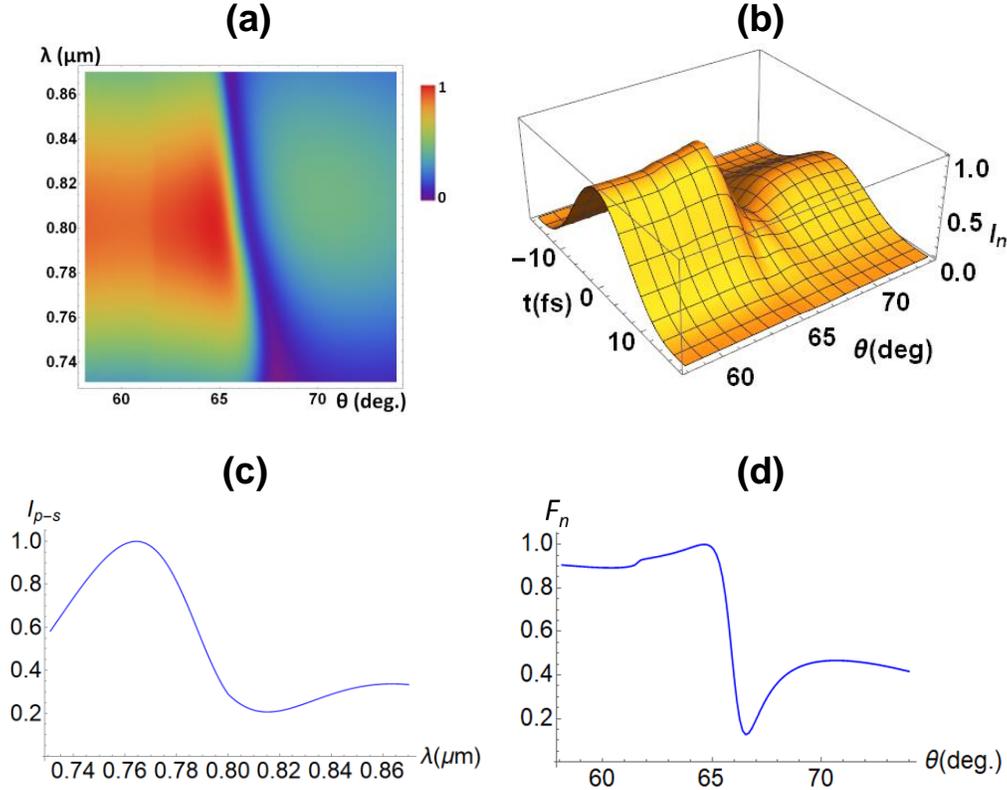

**Fig. 5.** Plots for a broadband pulse in a mixed polarization state: (a) A false color density plot of the normalized spectral intensity for different wavelengths λ and angles θ. (b) A 3D-plot of the normalized temporal intensity distribution of the reflected pulse at different incidence angles θ. (c) The spectral intensity distribution of the normalized reflected pulse at the angle $\theta=66.2°$. (d) The dependence of the normalized total radiation flux on the incidence angle (SPR-curve).

Two modes of detection can be used for sensing. Sending the laser pulse at different incidence angles results in an angular dependence of the radiation flux with a dip, which experiences an angular shift when the refractive index of the liquid is changed (Fig. 6. (a)). For a broadband pulse a spectral mode of detection is also possible. Figure 6(b) shows the changes of the spectrum for a pulse incident at an angle $\theta=66.2°$ when the index of refraction of the liquid is increased from that of water (blue line) by $\Delta n=0.01$ (the resultant SPR curve is shown by the red line).

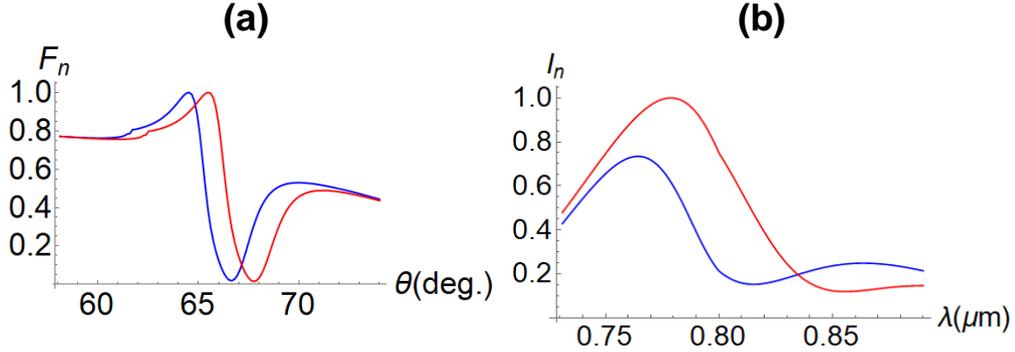

**Fig. 6.** The changes of the SPR dependences in response to changes of the refractive index of the liquid: (a) The angular dependence of the radiation flux of a pulse with mixed polarizations incident at different angles θ when the refractive index is equal to that of water $n_w$ (blue line) and when the liquid refractive index is incremented by Δn=0.01 (red line). (b) Spectral intensity distributions for a pulse with mixed polarizations when the refractive index is equal to that of water (blue line) and when this refractive index is incremented by Δn=0.01 (red line). In (a) and (b) the dependences are normalized to the maximum of the two curves.

To quantify the changes, the shift of the steep slope of the dependences can be determined. We define the angular slope position $\theta_{slope}$ (Fig. 6(a)) as the position of the middle point

$$\theta_{slope}= (\theta_{max}+ \theta_{min})/2, \qquad (12)$$

where $\theta_{max}$ and $\theta_{min}$ are the angular positions of the maximum and the minimum.

For the spectral response we define the slope position (Fig. 6(b)) as the wavelength position of the middle point

$$\lambda_{slope}= (\lambda_{max}+ \lambda_{min})/2, \qquad (13)$$

where $\lambda_{max}$ and $\lambda_{min}$ are the wavelengths corresponding to the maximum and the minimum.

Then we can calculate the response of the sensor, namely the displacement of the position of the steep slope of the SPR curve for a mixed polarization state in response to a change Δn of the liquid refractive index, $\Delta\theta=\theta_{slope}(n_w+\Delta n) - \theta_{slope}(n_w)$ (Fig. 7(a)) and $\Delta\lambda=\lambda_{slope}(n_w+\Delta n) - \lambda_{slope}(n_w)$ (Fig. 7(b)), which characterize the sensitivity of the two modes of measurements, angular and spectral respectively.

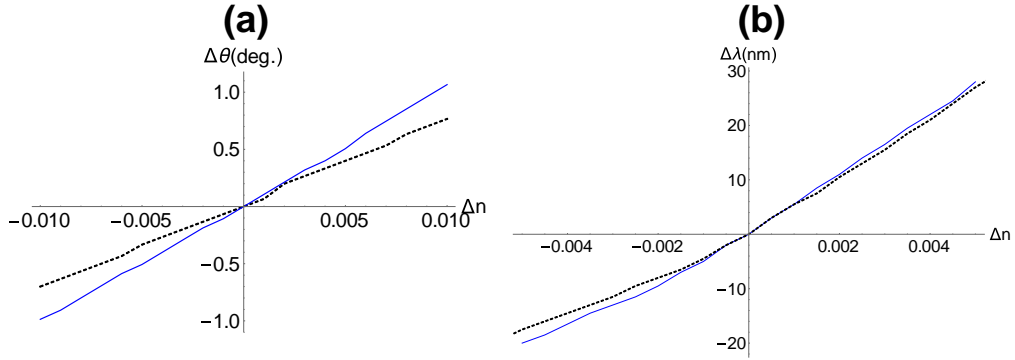

**Fig. 7.** The displacement of the position of the steep slope of the SPR curve for a mixed polarization state in response to a change Δn of the liquid refractive index: (a) angular shift Δθ and (b) wavelength shift Δλ.

To measure biomolecular interactions the gold surface can be functionalized by an additional (i.e. dextran) layer [40]. The presence of such a layer can somewhat reduce the response to the changes of the refractive index, since the latter will take place at a larger distance from the metal surface, where the SP field experiences additional decay. The effect of this additional layer (now 3$^{rd}$ medium) can be easily taken into account with the developed model, where the liquid is now the 4$^{th}$ medium. Assuming typical thickness of the dextran layer $d_3$=100nm and the refractive index $n_3$=1.45, we have calculated the response dependences, which are shown in Fig. 7(a,b) by the black dashed lines. Indeed, the response is reduced by about ~20% for the angular and by 7-15% for the spectral modes of measurements, respectively.

## 4. Conclusion

We have considered the characteristics of sensing with SPs by using coherent broadband laser pulses in the Kretschmann configuration. The intensity and phase dependences with the variations of the incidence angle and wavelength were calculated for the light components with polarizations parallel (interacting with SPs) and perpendicular (non-interacting) to the plane of incidence. By mixing these components the phase changes can be converted into intensity variations. For a pulse with mixed polarizations the phase shift in the vicinity of the resonance results in a steep intensity slope with a strongly asymmetric shape, characteristic of a Fano resonance. The degree of asymmetry can be varied by mixing p- and s-polarization components in different proportions.

It was shown that both angular and wavelength shifts of the intensity distribution can be used for sensing of the changes of the refractive index of the medium adjacent to the supporting SPs metal film. The angular and wavelength responses to refractive index changes of the adjacent medium were also simulated, when an additional dextran layer functionalizing the gold film was added to the system.

**Acknowledgments**
This work was supported by the Robert A. Welch Foundation, grant No. A1546 and the Qatar Foundation, grant NPRP 8-735-1-154.